\documentclass[12pt,authoryear]{elsarticle}

\usepackage{amssymb}
\usepackage{amsmath}
\usepackage{filecontents}
\usepackage{textcomp}
\usepackage{url}
\usepackage{lineno}
\usepackage{bm}
\usepackage{siunitx}
\usepackage{xcolor}
\usepackage{subfigure}

\journal{Earth and Planetary Science Letters}
\usepackage{setspace}
\begin{document}

\begin{frontmatter}

\title{Ohmic and viscous damping of inner core translational oscillations} 
\author[cnes,uga,cor1]{Paolo Personnettaz}
\author[uga]{David C\'ebron}
\author[uga]{Nathana\"el Schaeffer}
\author[uga]{Renaud Deguen}
\author[cnes]{Mioara Mandea}
\affiliation[cnes]{CNES - Centre National d'Etudes Spatiales, 2 Place Maurice Quentin, 75039, Paris, France}
\affiliation[uga]{Univ. Grenoble Alpes, Univ. Savoie Mont Blanc, CNRS, IRD, Univ. Gustave Eiffel, ISTerre, 38000 Grenoble, France}
\cortext[cor1]{paolo.personnettaz@univ-grenoble-alpes.fr}

\begin{highlights}
\item Slichter modes’ viscous dissipation is negligible, decaying over few centuries.
\item Three magnetohydrodynamic regimes exist: diffusive, skin layer, and Alfvén wave emission.
\item Earth’s Slichter modes lie between the skin layer and Alfvén wave emission regimes.
\item Ohmic damping occurs over 3–16 years, comparable to anelastic effects.
\item Equatorial modes decay at least twice as faster as the polar mode.
\end{highlights}

\begin{abstract}
Large earthquakes can trigger translational oscillations of Earth’s inner core (Slichter modes), yet their damping remains uncertain.
Using simulations, we quantify viscous and Ohmic dissipation in the fluid outer core.
Earth’s rotation splits the motion into one polar and two equatorial modes.
We explore all three and derive scaling laws for the quality factor with each dissipation mechanism. Viscous effects are negligible, confined to a thin layer at the inner core boundary. Ohmic dissipation dominates, with decay times of 3–16 years. Equatorial modes damp at least twice as fast as the polar mode. Our results suggest that Slichter modes can persist for years. Their continued non-detection is therefore more likely due to weak excitation than rapid damping.
\end{abstract}
\end{frontmatter}

\section{Introduction}

Understanding the dynamics of planetary interiors is central to interpreting their thermal evolution, magnetic field generation, and seismic response. In terrestrial planets and icy moons alike, the presence of a solid inner region surrounded by a liquid layer allows for global-scale translational oscillations. These co-called Slichter modes can be triggered when energetic events, such as impacts or large earthquakes, overcome gravitational energy. Detecting and quantifying such motions is crucial for better understanding planetary interiors' physical properties, including solid and liquid phase densities.

Despite decades of work, these modes have not yet been unambiguously detected. This raises a fundamental question: are Slichter modes intrinsically difficult to excite, or are they rapidly damped by physical processes within the core? Addressing this question is essential not only for Earth seismology, but also for extending normal mode studies to other planetary bodies.

In the case of the Earth, the inner core is free to weakly oscillate inside the liquid outer core. These translational oscillations correspond to a degree-one spheroidal mode ($l = 1$).
The Earth's rotation splits the oscillation into one polar mode ($m = 0$) along the rotation axis and two modes ($m = \pm 1$) in the equatorial plane.
The latter can be interpreted as an orbital motion of the inner core center of mass, one retrograde and the other prograde with respect to the Earth's rotation \citep{rosat2011review}.

Detection of translational oscillations were first suggested by \citet{slichter1961fundamental} following the 1960 Valdivia (Chile) earthquake of magnitude 9.6, leading to the designation of Slichter modes. 
This interpretation was later challenged by \citet{alsop1963free}, who questioned their detectability by gravimetry or strain seismography.
The reported \SI{86}{\min} period was inconsistent with the Bullen B density model \citep{bullen1963index} and lacked frequency splitting. \citet{won1973oscillation} incorporated viscous stresses and revised the period to \SI{7}{\hour} using the density model of \citet{birch1964density}, but neglected Coriolis and magnetic effects. \citet{busse1974free} subsequently analyzed the polar mode including rotation and spherical confinement, deriving an inviscid solution and showing that viscous and magnetic forces have little influence on the period, with only minor corrections from rotational and geometric effects.

\citet{smith1976translational} applied normal mode theory including rotation, elasticity, and a compressible, stratified outer core with elliptical boundaries, estimating a triplet period of \SI{4.5}{\hour}. Subsequent studies using different approaches obtained consistent results \citep{dahlen1979rotational,crossley1992slichter}. \citet{rieutord2002slichter} solved the coupled mechanical–hydrodynamic problem as an eigenvalue system, showing that the oscillation frequency is governed by gravitational and Coriolis forces, with minor contributions from viscous and pressure coupling at the ICB. Viscosity was found to have negligible influence on mode splitting, and this work remains the most comprehensive linear model, including various density profiles. Mantle motion was later incorporated by \citet{smylie1998viscous,smylie2000the}, with limited impact for Earth but potentially significant for other bodies such as Mercury \citep{grinfeld2005motion}. These developments have motivated further studies of translational oscillations in planets and moons \citep{escapa2011free,coyette2014slichter}.

As the inner core oscillates, it crosses isotherms, potentially inducing melting and solidification. \citet{wu1994gravity} neglected these effects due to the small oscillation amplitude, whereas \citet{peng1997effects} suggested that a km-thick mushy layer at the ICB could modify the period by \SI{0.5}{\percent}, and \citet{grinfeld2010effect} proposed a larger reduction based on a simplified model. Consistent with \citet{wu1994gravity}, we find phase change effects to be negligible for mode evolution. An upward displacement of the ICB leads to decompression cooling; if rapid, the process follows an adiabatic path that is less steep than the solidification gradient, causing local melting. The associated latent heat release restores thermal equilibrium. Quantitative estimates (Supplementary Material S1) show that only a tiny fraction ($\sim 10^{-10}$) of the topography induced by a 5 mm displacement is involved.

Overall, previous studies show that oscillation periods are primarily controlled by Archimedean forces and thus depend mainly on the density structure of Earth’s interior \cite[see the review of][]{rosat2011review}. In contrast to periods, dissipation mechanisms remain poorly constrained, with three main contributors identified: anelasticity of the solid layers, viscous shear in the liquid core, and Ohmic dissipation in the conducting regions. \citet{crossley1991excitation} examined damping from seismic anelasticity in the inner core and mantle, identifying the first as the primary dissipation site for Slichter modes.
Assuming that the PREM seismic model's quality factors can be applied to core periods, they estimated a quality factor of \num{5.2e3}, equivalent to a decay time of ~420 days. However, recent studies \cite[e.g.][]{lau2019anelasticity} challenge this assumption, though they primarily focus on the mantle.
In the following, we focus only on the remaining two dissipation sources, which are directly linked to the liquid and conductive properties of the outer core.

To estimate viscous dissipation, \citet{won1973oscillation} applied the analytical drag model for a sphere oscillating in an unbounded viscous fluid \citep{landau1987fluidmechanics}, obtaining $Q_\nu \sim \num{4e6}$–\num{4e7} for viscosities between \num{e-3} and \qty{e-1}{\pascal\second}. \citet{smith1976translational} derived a similar value ($Q_\nu \sim \num{3e6}$) from oscillating boundary layer theory \citep{batchelor1967introduction}. In contrast, \citet{smylie1992inner} proposed $Q_\nu \sim 100$ to explain observed splitting, implying an unrealistically large outer core viscosity of \qty{7.7e3}{\square\meter\per\second}. Analytical expressions accounting for rotation and confinement were later derived \citep{smylie1998viscous}, and used to infer viscosities up to \qty{e7}{\square\meter\per\second} near the inner core \citep{smylie1999viscosity}. These high values were subsequently questioned by \citet{rieutord2002slichter}, who argued they are non-physical and invalidate the associated interpretation.

Most studies neglected Ohmic dissipation \citep{won1973oscillation,crossley1991excitation}. \citet{busse1974free}, following \citet{toomre1974diurnal}, suggested that magnetic dissipation could exceed viscous effects, but without quantitative estimates. In contrast, \citet{smylie1992inner} argued for negligible magnetic damping, proposing a large quality factor $Q_\eta \sim \num{e10}$ based on the short oscillation period and thin skin depth. \citet{buffett1995magnetic}, noted BG95 hereinafter, provided an analytical boundary-layer model for magnetic damping, focusing on the polar mode and neglecting viscosity, rotation, and confinement. They identified two regimes controlled by the radial magnetic field $B_r$, with a transition from diffusion to Alfv\'en wave emission near $B_r=$~\SI{2}{\milli\tesla} at the ICB. For Earth, they estimated $Q_\eta$ between \num{2.2e3} and \num{5.8e5} for $B_r=$~\SI{10}{\milli\tesla} and \SI{0.5}{\milli\tesla}, respectively.

Beyond the Earth, translational oscillations are expected to occur in a wide range of planetary bodies, including Mercury and icy satellites hosting subsurface oceans. In these environments, the relative importance of viscous and magnetic dissipation may differ substantially due to variations in electrical conductivity, magnetic field strength, and layer thickness. Scaling laws are thus required to provide a framework for predicting the detectability and lifetime of such modes across planetary settings.

There remains a lack of consensus regarding the magnitude of various dissipation mechanisms.
In this study, we compute both viscous and Ohmic dissipation using simulations. Section \ref{sec:probdef} introduces the physical model used, followed by a description of its numerical implementation. For the regimes identified, we propose scaling laws for viscous and Ohmic dissipation in section \ref{sec:resu}.
These results are framed with the earlier theoretical formulation by BG95 and \citet{smylie1998viscous}, as well as the work of \citet{cebron2025hydromagnetic}. In the section \ref{sec:discCON}, we present updated estimates of the viscous and Ohmic quality factors for the Earth's inner core.

\section{Problem definition and methods} \label{sec:probdef}
In the rotating reference frame of the solid mantle, we simulate inner core translational oscillations using a magnetohydrodynamic model in which the solid inner core and liquid outer core are treated within a continuum framework, neglecting deformation of the solid regions. We discard viscous deformation in the inner core, justified by our assumption of an inner core viscosity exceeding \SI{2e16}{\pascal\second} \cite[consistent with][see Supplementary Material S5 for further discussion]{ritterbex2020viscosity}.
We neglect the thermosolutal transport and the corresponding buoyancy forces, assuming a well-mixed outer core. This simplification is justified by time scales separation, as the convective turnover time ($\approx$130 y) is five orders of magnitude longer than the oscillation period \citep{aubert2021interplay}.
Modeling the interaction of translational oscillations with dynamo generation lies beyond the scope of this study.
Instead, we impose an axial magnetic field, as it approximates the relevant feature of the Earth's magnetic field.

\subsection{Physical model}
\begin{figure}
   \centering
    \subfigure{
    \includegraphics[width=0.5\linewidth]{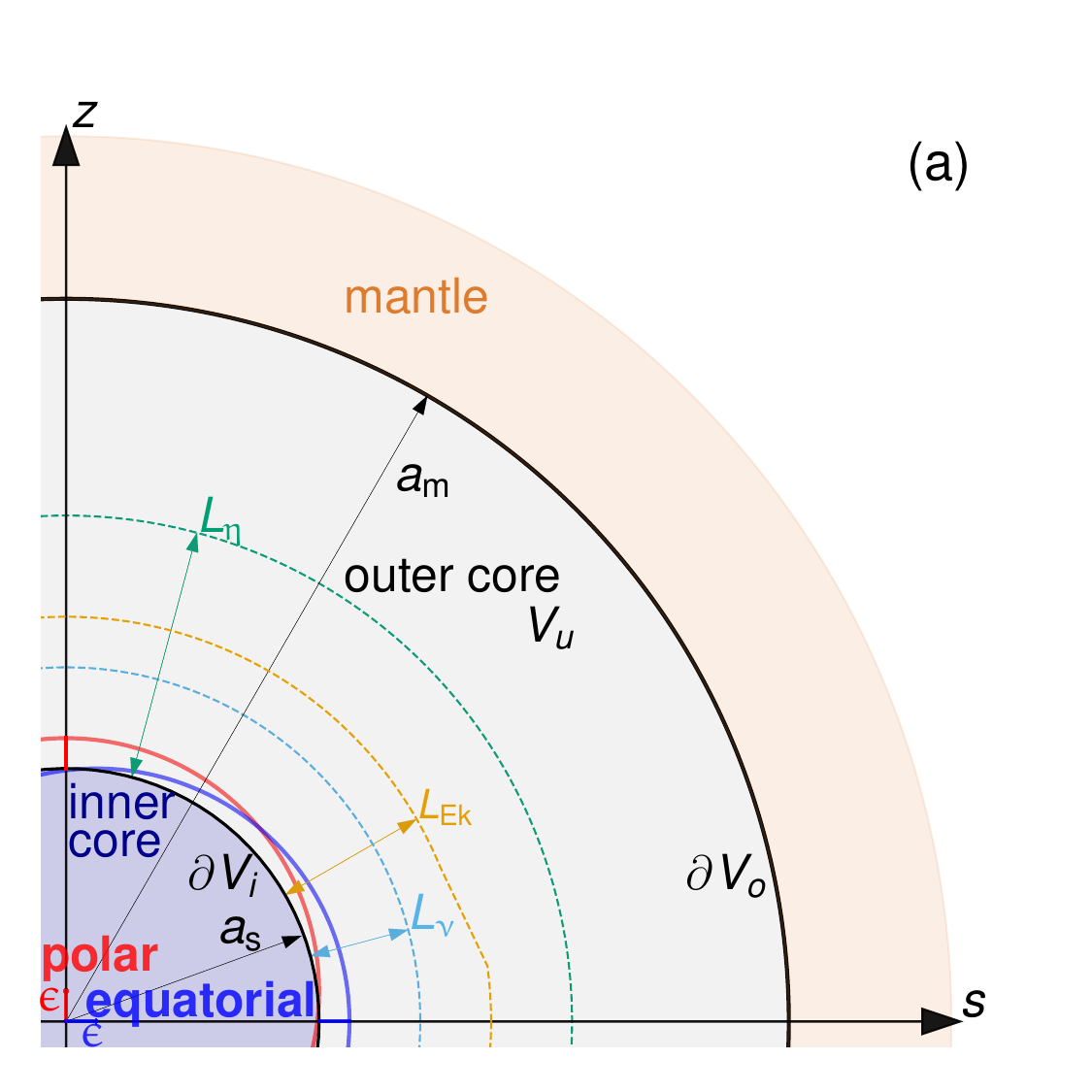}
    }
    \subfigure{
    \includegraphics[width=1.0\linewidth]{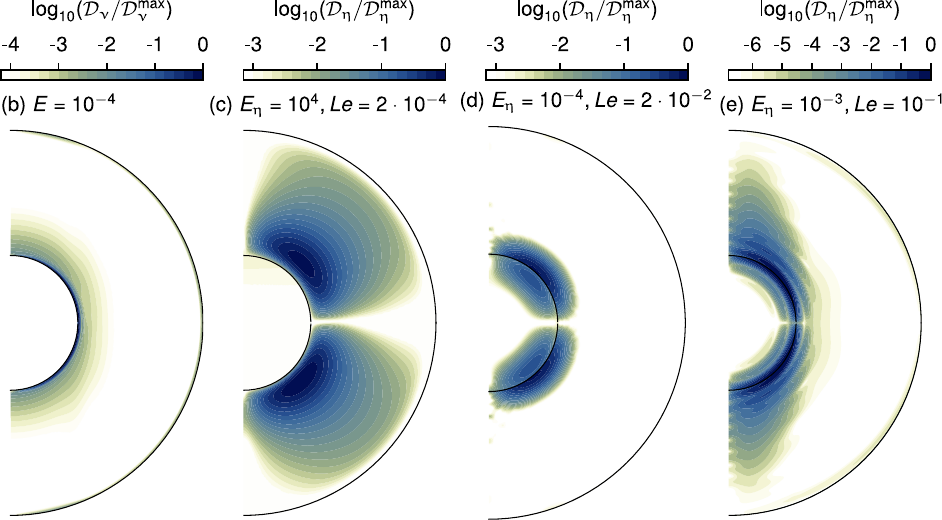}
    }
    \caption{(a) Sketch of a half meridional cut of the geophysical setup. (b) Viscous and (c,d,e) Ohmic dissipation spacial distribution on a meridional cut for polar forcing. (c) Diffusive regime, (d) skin layer regime, (e) Alfv\'en waves regime. Local dissipation is computed using the integrand of Eqs.~\ref{eq:viscdissipation} and  \ref{eq:Ohmdissipation}.}
    \label{fig:dissipationdistribution}
\end{figure}

We consider a conducting, viscous, incompressible fluid confined in a spherical shell of outer radius $a_\mathrm{m}$, driven by translational oscillations (frequency $\omega_\mathrm{s}$) of a solid inner sphere of radius $a_\mathrm{s}$ and density $\rho_\mathrm{s}$ (Fig.~\ref{fig:dissipationdistribution}(a)). The fluid has uniform density $\rho_\mathrm{f}$ and kinematic viscosity $\nu$. Fluid and solid inner domains share the same electrical conductivity $\sigma$ and magnetic permeability $\mu_0$, resulting in a uniform magnetic diffusivity $\eta$.

The system rotates at $\Omega_\mathrm{o} \bm{e_z}$ and is permeated by a uniform axial magnetic field $B_0\bm{e_z}$. Using $a_\mathrm{m}$ as length scale and $\Omega_\mathrm{o}^{-1}$ as time scale, the governing equations are (in the frame of reference rotating at $\Omega_\mathrm{o} \bm{e_z}$)
\begin{equation}
	\partial_t\bm{u}+\bm{u}\cdot\nabla\bm{u} + 2\bm{e_z}\times\bm{u} = -\nabla p + E \nabla^2\bm{u} + Le^2 (\nabla \times \bm{b}) \times \bm{b} , \ \ \nabla \cdot \bm{u}  = 0,
 \label{eq:momentum}
\end{equation}
in the fluid domain $V_{u}$ of dimensionless volume $4\pi(1-a^3)/3$, with inner-outer radii ratio $a=a_\mathrm{s}/a_\mathrm{m}$; $\bm{u}$ is the velocity, $p$ the reduced pressure (including centrifugal effects), and $\bm{b}$ the magnetic field. The Ekman and Lehnert numbers are $E=\nu \Omega_\mathrm{o}^{-1} a_\mathrm{m}^{-2}$ and $Le = B_0(\sqrt{\rho_\mathrm{f}\mu_0}\Omega_\mathrm{o} a_\mathrm{m})^{-1}$.
The magnetic field evolves in the whole space following the induction equation and Gauss's law, respectively
\begin{equation}
	\partial_t\bm{b}= \nabla\times(\bm{u}\times\bm{b}) + E_\eta \nabla^2\bm{b}   , \ \ \nabla \cdot \bm{b}  = 0
 \label{eq:induction}
\end{equation}
where $Pm =\nu \eta^{-1}$ is the magnetic Prandtl number, and $E_\eta=E Pm^{-1}$ is the magnetic Ekman number.
In our model, the induction term $\nabla\times(\bm{u}\times\bm{b})$ is non-zero solely within the fluid domain $V_u$.
Neglecting induction effects in the solid domains is valid only at leading order.
The region surrounding the fluid shell is treated as a perfect electrical insulator, with magnetic boundary conditions as in \cite{christensen}. At the interface between the fluid and the inner solid domain, we impose continuity of the magnetic field and electrical currents.

Rather than physically displacing the inner sphere, translational oscillations are modeled by prescribing a time-varying velocity at the inner boundary $\partial V_\mathrm{i}$ that mimics the expected dynamics \cite[following][]{rieutord2002slichter}.
The cylindrical coordinate system (cylindrical radius $s$, axial coordinate $z$, and azimuthal angle $\varphi$) yields a more compact formulation for these modes.
For polar mode, only the axial velocity $u_z$ evolves as
\begin{equation}
     u_s = 0  , \ u_z = \epsilon\omega \cos(\omega t), \  u_\varphi = 0 \  \mathrm{at} \ \partial V_\mathrm{i}  .
     \label{eq:polar}
\end{equation}
where $\omega$ is the dimensionless forcing frequency and $\epsilon$ is the dimensionless  displacement amplitude of the oscillation.
For equatorial modes, the cylindrical radial velocity $u_s$ and the azimuthal velocity $u_\varphi$ are imposed as
\begin{equation}
    u_s^{\pm} = \epsilon\omega \cos(\omega t  \pm\phi) ,  \ u_z^{\pm} = 0, \  u_\varphi^{\pm} = \mp\epsilon\omega \sin(\omega t \pm\phi) \  \mathrm{at} \ \partial V_\mathrm{i} ,
    \label{eq:equatorial}
\end{equation}
where the plus and minus signs correspond to the retrograde and prograde modes, respectively, following BG95 convention.
These approximate boundary conditions are appropriate when the oscillation is limited within the viscous boundary layer, as discussed in supplementary material S2.
We assume a no-slip boundary condition $\bm{u}=\bm{0}$ at $r=a_\mathrm{m}$.

To understand the response to translational oscillations, we devise diagnostic quantities.
Time-averaged dimensionless viscous and Ohmic dissipation are  \citep{lin2018tidal,lin2020ohmic}
\begin{eqnarray}
    \mathcal{D}_\nu &=& \langle E\int_{V_u} ( [\nabla \bm{u} +\nabla \bm{u}^\top  ]\boldsymbol{:}\nabla\bm{u}) \mathrm{d}V \rangle_t  , \label{eq:viscdissipation} \\ \mathcal{D}_\eta &=& \langle  Le^2 E Pm^{-1} \int_{V_b}  |\nabla \times \bm{b}|^ 2 \mathrm{d}V \rangle_t\ . \label{eq:Ohmdissipation}
\end{eqnarray}
To convert them into physical units, it is sufficient to multiply by the power scale $\rho_\mathrm{f} \Omega_\mathrm{o}^3 a_\mathrm{m}^5$.
The quality factors of the two dissipation mechanisms are $ Q_j = E_\mathrm{kin}^\mathrm{max} \omega  \mathcal{D}_j^{-1}$ where $j \in (\nu, \eta)$, with the maximum kinetic energy of the oscillating solid sphere
\begin{equation}
    E_\mathrm{kin}^\mathrm{max} =\frac{1}{2} \frac{M_\mathrm{s}}{\rho_\mathrm{f} a_\mathrm{m}^3} (\epsilon\omega)^2= \frac{4\pi}{6} \frac{ \rho_\mathrm{s} a_\mathrm{s}^3}{\rho_\mathrm{f} a_\mathrm{m} ^3} (\epsilon\omega)^2 \ = \chi a^3 (\epsilon\omega)^2 ,
    \label{eq:kineticEnergy}
\end{equation}
where $M_\mathrm{s}$ is the inner sphere mass, and $\chi=2\pi\rho_\mathrm{s}/(3\rho_\mathrm{f})$ is the kinetic energy prefactor.
Note that while the two modes share the same maximum kinetic energy, the time-averaged kinetic energy of the equatorial mode is twice that of the polar mode over a period. This follows directly from applying the Pythagorean trigonometric identity to the equatorial inner core motion (see Eq.~\ref{eq:equatorial}), revealing that the instantaneous kinetic energy of the equatorial mode remains constant, whereas that of the polar mode varies as $\propto\cos^2(\omega t)$. Following \citet{crossley1991excitation}, the e-folding damping time $\tau_j$ is defined as $\tau_j=2 Q_j/\omega_\mathrm{s}$, with $j \in (\nu, \eta)$.

\subsection{Numerical implementation}
For small oscillation amplitudes, the problem is spherically symmetric, allowing the use of spherical harmonics as basis functions \cite[as e.g. done by][]{rieutord2002slichter}. We integrate the evolution equations~\ref{eq:momentum} and \ref{eq:induction} using the finite-difference and pseudospectral code XSHELLS \citep{schaeffer2017turbulent}.

A poloidal-toroidal decomposition is used to represent the velocity and magnetic fields at each radial point, $\bm{u} = \nabla \times (\mathcal{T} \bm{r}) +  \nabla \times \nabla \times (\mathcal{P} \bm{r})$.
The toroidal $\mathcal{T}$ and poloidal $\mathcal{P}$ scalar fields are described by a linear combination of spherical harmonics $Y_{m}^{l}$ with a maximum degree of $l = l_\mathrm{max}$ and an azimuthal wavenumber of $m = m_\mathrm{max}$.
The shtns library, detailed in \citet{schaeffer2013efficient}, handles the spherical harmonic transforms.
Over the radius, XSHELLS makes use of a second-order finite difference scheme.
Time integration is performed using a third-order semi-implicit backward differentiation scheme SBDF3  \citep{ascher1995implicit}.
At $E=$~\num{e-9}, the radial resolution is of $1024$ nodes, enough points are reserved to resolve the boundary layers and the oscillation displacement $\epsilon$, close to the inner radius.
The spectral resolution is  $ l_\mathrm{max} = 512$ , and $ m_\mathrm{max} = 7$.
At the forcing amplitudes studied, the problem remains linear, and can be fully described by a few azimuthal wavenumber $m$.
The code had undergone extensive validations over the years \cite[e.g.][]{marti2014full}.

To study translational oscillations, new boundary conditions on velocity have been implemented (Eqs.  \ref{eq:polar} and \ref{eq:equatorial}).
The polar oscillation can be modeled by changing the projection ($l=1$, $m=0$) of the poloidal velocity field $\mathcal{P}_u$ and its radial derivative at the inner boundary.
Meanwhile, equatorial oscillations are described by the ($l=1$, $m=\pm1$) projection of the poloidal velocity field. The toroidal velocity field at the inner boundary is set to zero, meaning that there is no injection of angular momentum into the fluid.

For translational oscillations, both hydrodynamic and magnetohydrodynamic solutions were benchmarked against results from the finite-element software COMSOL Multiphysics\textsuperscript{\textregistered}. In this framework, the geometric displacement of the inner core is truly modeled using an arbitrary Lagrangian–Eulerian method that deforms the mesh. Although this approach allows arbitrary displacements, it is computationally expensive and unsuitable for extensive parameter surveys. The comparison is detailed in Supplementary Material S3. All results presented below are therefore obtained with the finite-difference spectral code XSHELLS, chosen for its computational efficiency.

\begin{table*}
	\centering
	\begin{tabular}{llll}
		\textbf{Parameter}                     & \textbf{Symbol}  & \textbf{Value}                            & \textbf{Reference}                            \\   \hline
		\text{Inner core radius}               & $a_\mathrm{s}$            & \SI{1.221e6}{\meter}                      & \citet{dziewonski1981preliminary}            \\
		\text{Outer core radius}               & $a_\mathrm{m}$            & \SI{3.480e6}{\meter}                      & \citet{dziewonski1981preliminary}            \\
		\text{Inner core density}              & $\rho_\mathrm{s}$          & \SI{1.3e4}{\kilogram\per\cubic\meter}     & \citet{dziewonski1981preliminary}            \\
		\text{Outer core density}              & $\rho_\mathrm{f}$          & \SI{1.2e4}{\kilogram\per\cubic\meter}     & \citet{dziewonski1981preliminary}            \\
		\text{Kinematic viscosity}             & $\nu$            & \SI{e-6}{\meter\squared\per\second}       & \citet{mineev2004viscosity}                \\
		\text{Magnetic diffusivity}            & $\eta$           & \SI{1}{\meter\squared\per\second}         & \citet{nataf2024dynamic}                                       \\
		\text{Spin-rate}                       & $\Omega_\mathrm{o}$        & \SI{7.29e-5}{\per\second}                 & \citet{groten2000report}                                             \\
		\text{Magnetic field at ICB}                  & $B_0$            & \SI{2e-3}{\tesla}                         & \citet{gillet2015planetary}                       \\ \hline
       	\text{Oscillation amplitude}           & $\epsilon_\mathrm{s}$       & \SI{0.005}{\meter}                         & \citet{rosat2011review}                      \\
        \text{Viscous skin depth}                    & $L_\nu$           & \SI{0.07}{\meter}                             &  assuming  $\omega_\mathrm{s}=$ \SI{4e-4}{\per\second}                               \\
        \text{Ekman depth}         & $L_E$           & \SI{0.12}{\meter}                              &                                        \\
        \text{Magnetic skin depth}                  & $L_\eta$           & \SI{70}{\meter}                             &   assuming  $\omega_\mathrm{s}=$\SI{4e-4}{\per\second}                                    \\
        \hline
		\text{Ekman Number}                    & $E$           & \SI{1.1e-15}{}                             &                                 \\
	    \text{Magnetic Prandtl Number}         & $Pm$           & \SI{e-6}{}                              &                                        \\
		\text{Lehnert Number}                  & $Le$           & \SI{6.4e-5}{}                             &                                       \\
		\text{Magnetic Ekman Number}           & $E_\eta$       & \SI{1.1e-9}{}                              &    \\

		\text{Inner-outer core ratio}           & $a$       & 0.35                          &

		\\ \hline
	\end{tabular}
	\caption{Earth's parameters. ICB is the inner core boundary.}
	\label{tab:parameters}
\end{table*}

\subsection{Earth's parameters and simulations targets}
\label{sec:Earthparameters}
In this section, we define the Earth's values, summarized in Tab.~\ref{tab:parameters}, and the simulation parameters explored.
The Earth's inner core average density is $ \rho_\mathrm{s} =$~\qty{1.3e4}{\kilo\gram\per\cubic\meter} and its radius is $a_\mathrm{s}=$~\qty{1221}{\kilo\meter}.
The outer core fluid density is $\rho_\mathrm{f}=\qty{1.2e4}{\kilogram\per\cubic\meter}$ and its radius is $a_\mathrm{m}=$~\qty{3480}{\kilo\meter}, see \citep{dziewonski1981preliminary}.
The resulting kinetic energy prefactor (Eq.~\ref{eq:kineticEnergy}) is $\chi=4\pi\rho_\mathrm{s}/(6\rho_\mathrm{f})=2.27$.

We expect $\omega$ to vary between 3.8 and 9.2 \citep{rosat2011review}.
For the final estimation, we adopt the frequencies reported in Table 1 of \citet{rieutord2002slichter} computed with the homogeneous density each layer (as in Tab.~\ref{tab:parameters}), and validated against \citet{busse1974free}'s model.
Resulting in $\omega=$5.73 for the polar mode, 6.26 and 5.20 for the retrograde and prograde equatorial modes, respectively.
We performed simulations with $\omega$ spanning from 2.5 to 12, to identify the role of the frequency in the dissipation scaling.
Considering $\omega>2$ allows us to avoid the inertial wave regime \citep{goertler1944einige}.

Estimates for the outer core kinematic viscosity range between \qty{3e-7}{\square\meter\per\second} and \qty{5e-6}{\square\meter\per\second} \citep{mineev2004viscosity,landeau2022sustaining}.
We assume here the intermediate value of \qty{e-6}{\square\meter\per\second}, which gives $E=$\num{1.1e-15}.
We could not reach this parameter directly with our simulations.
Hence, we establish scaling laws based on reachable parameters, up to $E$=\num{e-9}, to extrapolate the dissipation to Earth's conditions.

According to \citet{rosat2020observations}, the 1960 Valdivia earthquake resulted in a inner core 
displacement of \qty{5}{\milli\meter}.
We adopt this value for estimating the dissipation in the Earth's condition. As sketched in Figure~\ref{fig:dissipationdistribution}(a), these oscillations are thus buried inside the Ekman boundary layer of thickness $L_{E}=\sqrt{\nu/\Omega_\mathrm{o}}\approx$~\qty{12}{\centi\meter}, and in viscous skin layer of thickness $L_\mathrm{\nu}= \sqrt{2 \nu/\omega_\mathrm{s}}\approx$~\qty{7}{\centi\meter}.
To comply with physical constraints, we restricted the displacement amplitude to be below these two length scales.
From the viscous skin depth, we can define the oscillation Reynolds number $Re_\omega = \omega_\mathrm{s} a_\mathrm{s}^{2} \nu^{-1}$, the Earth value being approximately \num{6e14}.

The magnetohydrodynamic problem introduces two additional parameters: the imposed magnetic field strength and the magnetic diffusivity (assumed uniform). Direct measurements of the magnetic field at the ICB are unavailable. Estimates from inner core nutation studies suggest $B_r=\SIrange{9}{16}{\milli\tesla}$ \citep{koot2013role}, whereas torsional wave analyses indicate $\sim$\qty{2}{\milli\tesla} \citep{gillet2010fast,gillet2015planetary}. We adopt the latter value, consistent with estimates based on dynamo simulations \citep{schaeffer2017turbulent} and the observed core–mantle boundary field of \qty{0.33}{\milli\tesla} \citep{gillet2015planetary}. This yields a Lehnert number $Le \approx \num{6.4e-5}$. We explore $Le$ in the range \num{2e-5}–\num{e-1}, corresponding to a regime where magnetic forces remain weaker than Coriolis forces, as in Earth’s core.

The electrical conductivity of iron under outer core conditions is $\sim$\qty{e6}{\siemens\per\meter}, although values remain debated \citep{ramakrishna2023electrical}. This implies a magnetic diffusivity $\eta \approx \qty{1}{\square\meter\per\second}$ and magnetic Prandtl number $Pm \approx \num{e-6}$ \citep{nataf2024dynamic}. For $Pm \ll 1$, the magnetic skin depth $L_\eta=\sqrt{2\eta\omega_\mathrm{s}^{-1}}$ exceeds the viscous one ($L_\eta>L_\nu$). For Earth, $L_\eta \approx \qty{70}{\meter}$, about $10^3$ times larger than $L_\nu$, so oscillations are confined within both viscous and magnetic boundary layers.

To preserve the relevant scale hierarchy, we consider $Pm$ in the range \num{e-6}–0.6. The magnetic Reynolds number based on the oscillation frequency is $Rm_{\omega} = Re_{\omega} Pm = 2 \left(a_\mathrm{s} \, L_\eta^{-1}\right)^{2}$, with an expected Earth value of $\sim$\num{6e8}. The large values of $Re_\omega$ and $Rm_\omega$ reflect the choice of the inner core radius as length scale. By varying $E$ and $Pm$, we explore $Rm_\omega$ from \num{e-5} to \num{e7}.

Following BG95, we expect two regimes as a function of magnetic field strength: a diffusive regime and an Alfv\'en wave regime. The magnetic perturbation induced by the oscillation is expected to vary over the magnetic skin depth. These considerations motivate the introduction of a dimensionless number comparing the diffusive timescale to the Alfv\'en wave period: the skin-depth Lundquist number
\begin{equation}
    Lu_\omega= \frac{v_A L_\eta}{\sqrt{2}\eta}= \frac{Le}{\sqrt{E_\eta\omega}}
\end{equation}
where $v_A=B_0(\rho\mu)^{-1/2}$ is the Alfv\'en speed.
BG95 used a parameter named $\Lambda$, which is $\Lambda=Lu_\omega^2$.
For the Earth's translational oscillation, $Lu_\omega \approx0.8$ with the ICB values of table \ref{tab:parameters}.
The Earth's case is therefore at the transition between the diffusive regime and an Alfvén wave dominated regime.
We characterize both regimes, exploring $Lu_\omega$ between \num{e-7} and \num{20}.

\section{Results} \label{sec:resu}

\subsection{Viscous dissipation}
Viscous shear dissipates kinetic energy as heat, primarily within the inner hydrodynamic boundary layer where velocity gradients are largest. This is illustrated in Fig.~\ref{fig:dissipationdistribution}(b) for a polar-mode simulation at $E=\num{e-4}$, where the viscous layer is thickened relative to Earth conditions.

\begin{figure*}
    \centering
    \subfigure{
    \includegraphics[width=0.7\linewidth]{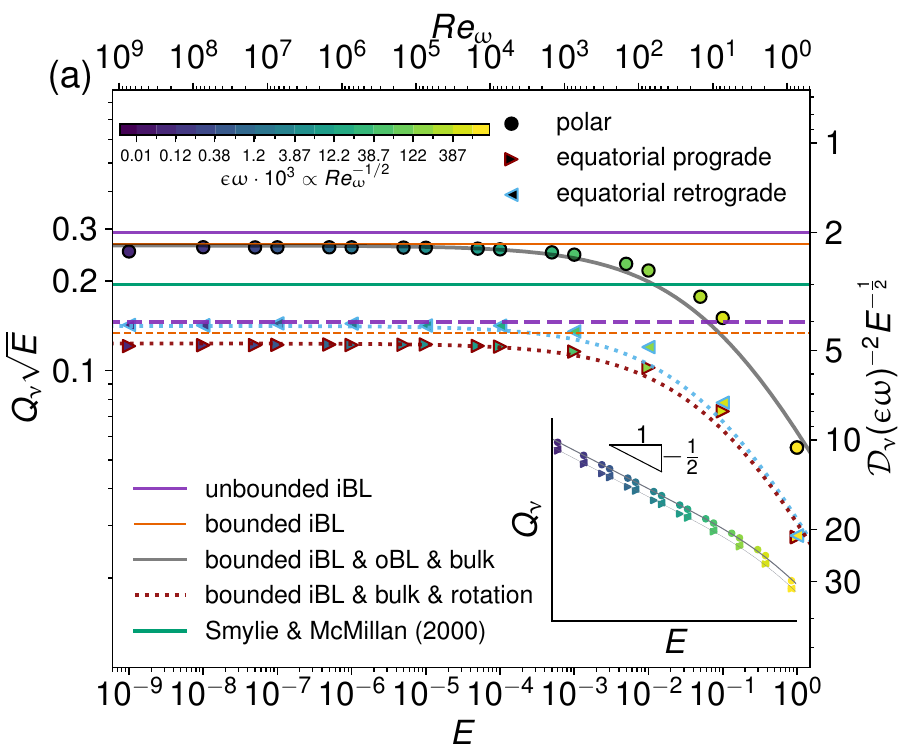}
    \label{fig:QnuEk}
    }
    \subfigure{
    \includegraphics[width=1.0\linewidth]{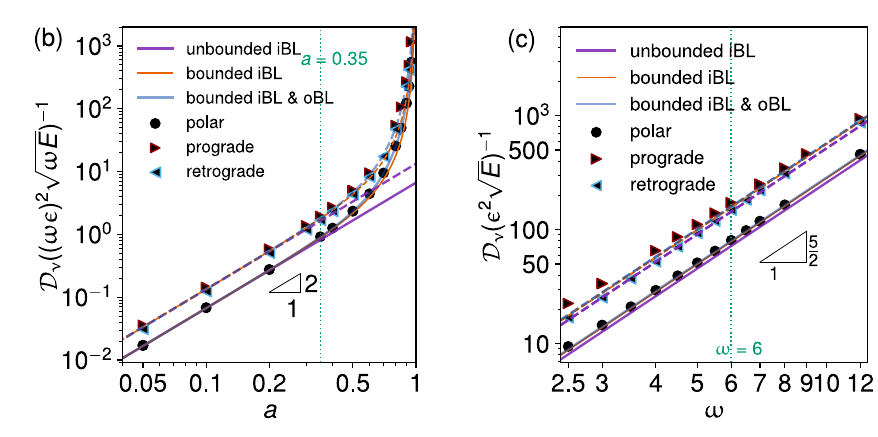}
    \label{fig:Dnuradii}
    }
    \caption{(a) Corrected viscous quality factor (left) and dissipation (right) vs. Ekman (bottom) and oscillation Reynolds number (top). (a) inset: viscous quality factor vs. Ekman number.
    Scaled viscous dissipation vs. inner-outer radii ratio (b) and dimensionless oscillation frequency (c). Simulations (scattered points). Polar-theory, Eq.~\ref{eq:viscousDissipation}, (solid line), \texttt{unbounded iBL}: $d_\nu=3\pi/\sqrt{2}$, \texttt{bounded iBL}: $d_\nu=3\pi/\sqrt{2}(1-a^3)^{-2}$, \texttt{bounded iBL \& oBL}: Eq.~\ref{eq:Dnu_fa}, \texttt{bounded iBL \& oBL \& bulk}: Eq.~\ref{eq:Dnu_fa} and  Eq.~\ref{eq:viscousDissipationVolume}, \cite{smylie2000the}: $d_\nu=\pi\sqrt{2}$. Equatorial-theory (dashed lines) as before with a multiplicative factor 2, except for \texttt{bounded iBL \& bulk \& rotation} (dotted line): Eq.~\ref{eq:Dnu_fa2}. \texttt{iBL}: inner and \texttt{oBL}: outer boundary layers.  Parameters: $a=0.35$ in (a,c), $\omega=6$ in (a,b).   Earth values (vertical dotted line)}
    \label{fig:Qnufactor}
\end{figure*}

To extrapolate to Earth's conditions, we investigate how the different parameters influence dissipation and quality factor.
The viscous quality factor $Q_\nu$ scales as the inverse square-root of the Ekman number consistently for several decades, as shown in the right-bottom inset in Figure~\ref{fig:Qnufactor}(a).
To compare it with different theoretical models, we normalize it with $\sqrt{E}$ in Figure~\ref{fig:Qnufactor}(a).
Polar and equatorial modes are presented as dots and triangles, respectively.
The two equatorial modes, prograde (red right-pointing triangles) and retrograde (azure left-pointing triangles), are characterized by slightly different $\mathcal{D}_\nu$ and $Q_\nu$ at equivalent parameters.
The prograde motion is the most dissipative one of the three.

Assuming power laws with simple exponents, our simulations show that the dimensionless viscous dissipation scales as (for $E \ll 1$ and $a \ll 1$)
\begin{equation}
    \mathcal{D}_\nu= d_\nu (\omega E)^{1/2} (\epsilon\omega)^2  a^2  = d_\nu Re_{\omega}^{-1/2} (\epsilon\omega)^2 \omega a^3
\label{eq:viscousDissipation}
\end{equation}
where $d_\nu$ is the numerical prefactor, equal to 7.5 for the polar mode, 16 and 14 for the prograde and retrograde equatorial modes, respectively. 
These numerical prefactors, and the following, are computed for Earth's parameters ($a=0.35$, $\omega=6$).
As a reference, the oscillation Reynolds number $Re_\omega$ is on the top x-axis of Figure~\ref{fig:Qnufactor}(a). 
Ignoring the prefactor, Eq.\ref{eq:viscousDissipation} yields to a dimensional dissipation independent of $a_\mathrm{m}$ and $\Omega_\mathrm{o}$.

The resulting quality factor is
\begin{equation}
Q_\nu= \chi d_\nu^{-1} E^{-1/2} \omega^{1/2} a = \chi d_\nu^{-1} Re_{\omega}^{1/2}  \ ,
\label{eq:viscousQualityFactor}
\end{equation}
which is independent of both velocity and displacement amplitude of the oscillations.
This is satisfied for an oscillation amplitude within the viscous layer. The dimensionless frequency should be larger than $2$ to avoid inertial waves \citep{goertler1944einige}.
Moreover, magneto-hydrodynamic phenomena are inconsequential for small $Pm$ and $Lu_\omega$,  especially when the inner and outer cores share similar electrical properties \citep{cebron2025hydromagnetic}.

The $\nu^{-1/2}$ scaling of viscous dissipation has been previously proposed \citep[e.g.][]{smith1976translational,toomre1974diurnal}, but without constraining its dependence on other parameters or the prefactor. Our scaling law (Eq.~\ref{eq:viscousDissipation}) is consistent with the classical solution for a sphere oscillating in an unbounded viscous fluid derived from Stokes theory \citep{stokes1851effect,batchelor1967introduction}, which accounts for dissipation within the inner boundary layer (\texttt{unbounded iBL}; violet line in Fig.~\ref{fig:Qnufactor}). In this limit, the analytical prefactor for the polar mode is $d_\nu=3\pi/\sqrt{2} \approx 6.66$, about \qty{13}{\percent} different from our numerical estimate. Considering only tangential viscous stress yields $d_\nu=\pi\sqrt{2} \approx 4.44$ \citep{smylie2000the} (green line in Fig.~\ref{fig:Qnufactor}(a)), corresponding to two-thirds of the total dissipation \citep[as expected, see][]{cebron2025hydromagnetic}.

Our results show that confinement significantly modifies viscous dissipation. For the polar mode, these effects are already implicitly hidden in the hydrodynamic solution of \citet{stokes1851effect} for a sphere oscillating in a bounded viscous fluid, and have largely been overlooked in previous studies. In the rapid oscillation limit, Stokes theory yields
\begin{equation}
    d^\mathrm{p}_\nu \approx \frac{3\pi}{\sqrt{2}} \frac{1+ a^4}{(1-a^3)^2} \ .
    \label{eq:Dnu_fa}
\end{equation}
where the $(1-a^3)^2$ term captures the effect of confinement at the core–mantle boundary and increases dissipation by \qty{9}{\percent} at $a=0.35$ (orange line in Fig.~\ref{fig:Qnufactor}(a), \texttt{bounded} \texttt{iBL}), while the $a^4$ term accounts for the outer boundary layer contribution (\qty{1}{\percent}). For Earth parameters, Eq.~\ref{eq:Dnu_fa} gives $d^\mathrm{p}_\nu=7.4$, within \qty{1}{\percent} of our numerical estimate, demonstrating excellent agreement.

Figure~\ref{fig:Qnufactor}(b) highlights the dependence on the radii ratio $a$. The unbounded solution (\texttt{unbounded} \texttt{iBL}, violet line; Eq.~\ref{eq:viscousDissipation} with $d_\nu=3\pi/\sqrt{2}$) reproduces the quadratic scaling at small $a$ but deviates at larger values. Earth lies near this transition. Including confinement via Eq.~\ref{eq:Dnu_fa} (\texttt{bounded} \texttt{iBL} \texttt{\&} \texttt{oBL}, grey line) restores agreement with simulations. In thin shells, confinement can enhance viscous dissipation by orders of magnitude. This scaling with fluid volume $(1-a^3)$ is further validated in Supplementary Material S4.

Outside the inertial-wave regime, viscous dissipation scales as $\omega^{5/2}$ (Fig.~\ref{fig:Qnufactor}(c)). In the fast oscillation limit, equatorial modes dissipate approximately twice as much as the polar mode, $d_\nu^{\pm}/d_\nu^\mathrm{p}\approx2$ \citep{cebron2025hydromagnetic}, because they inject twice the energy over one period for the same amplitude $\epsilon\omega$, although the maximum kinetic energy (Eq.~\ref{eq:kineticEnergy}) is identical.

Rotation effects are required to explain the difference between prograde and retrograde modes (Fig.~\ref{fig:Qnufactor}). From simulations, we refine the prefactor as
\begin{equation}
    d^\mathrm{\pm}_\nu \approx 2 d^\mathrm{p}_\nu \left(1 \mp \alpha \omega^{-1}\right) .
    \label{eq:Dnu_fa2}
\end{equation}
where $\alpha=0.5$ for $a=0.35$, and the $\pm$ signs denote retrograde and prograde modes. This correction reproduces the asymmetry between equatorial modes (azure and red dotted lines in Fig.~\ref{fig:Qnufactor}(a), \texttt{bounded iBL \& bulk \& rotation}). Neglecting higher-order terms ($\propto \omega^{-2}$) yields $(d_\nu^{+}+d_\nu^{-})/2=2d_\nu^\mathrm{p}$. The difference between the two equatorial modes vanishes in the absence of Coriolis effects, consistent with boundary-layer theory \citep{cebron2025hydromagnetic}.

At large $E$, viscous dissipation in the irrotational bulk flow becomes significant and dominates the drag of a rapidly oscillating sphere with stress-free boundaries. Extending \citet{levich1949motion} to bounded domains, \citet{cebron2025hydromagnetic} showed that this contribution is
\begin{equation}
    \mathcal{D}^\mathrm{bulk}_{\nu} = 6\pi \frac{1-a^5}{(1-a^3)^2}  a E(\epsilon\omega)^2 = 6\pi \frac{1-a^5}{(1-a^3)^2}\frac{a^3\omega}{Re_\omega} (\epsilon\omega)^2 \ ,
    \label{eq:viscousDissipationVolume}
\end{equation}
which scales linearly with viscosity (neglecting rotation). Including this bulk term alongside boundary dissipation (Eq.~\ref{eq:viscousDissipation}) reproduces the large-$E$ behavior observed in simulations (gray curve, Fig.~\ref{fig:Qnufactor}(a), \texttt{bounded} \texttt{iBL} \texttt{\&} \texttt{oBL} \texttt{\&} \texttt{bulk}). Although negligible for Earth-like viscosities, this contribution may be relevant in studies assuming higher viscosities, such as \citet{smylie1999viscosity}. Estimates for Earth conditions are discussed below.

\subsection{Ohmic dissipation}
\begin{figure*}
   \begin{center}
    \subfigure{
    \includegraphics[width=1.1\linewidth]{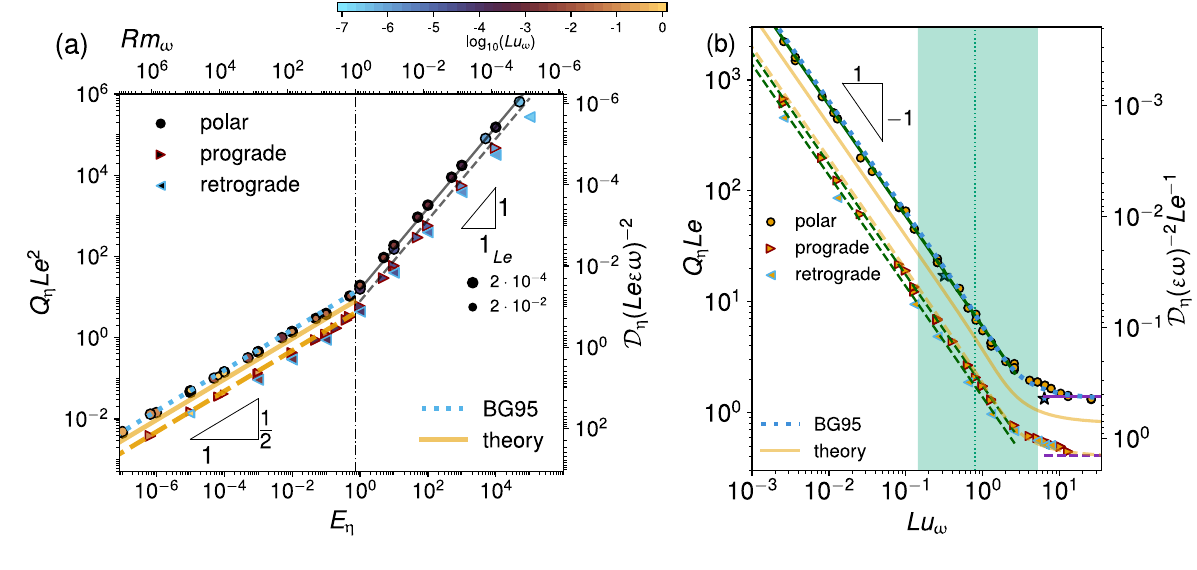}
    }
    \subfigure{
    \includegraphics[width=1.1\linewidth]{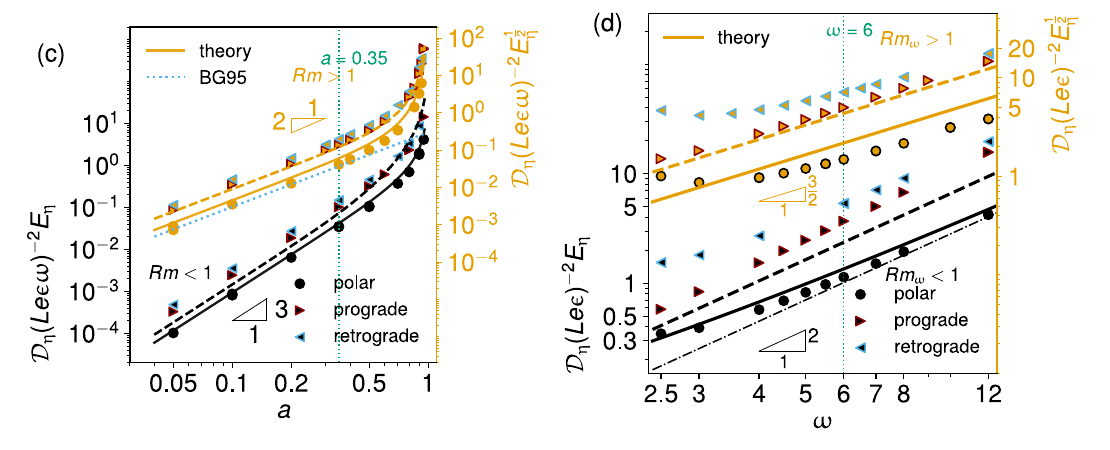}
    }
    \caption{(a) Corrected Ohmic quality factor (left) and dissipation (right) vs. magnetic Ekman number (bottom) and oscillation-based magnetic Reynolds number (top), for polar and equatorial (prograde and retrograde) modes; colors indicate the skin-depth Lundquist number ($Lu_\omega < 1$). (b) Same quantities versus $Lu_\omega$ ($Rm_\omega > 1$).  Expected $Lu_\omega$ for Earth (green area). (c,d) Corrected Ohmic dissipation vs. inner–outer radii ratio and dimensionless frequency. Black lines correspond to Eq.~\ref{eq:ohmicDissipationVolume} with analytical prefactor, and orange lines to Eq.~\ref{eq:Qeta4} with $d_{\eta,2}=d^{\star}_{\eta,2}$  and $\beta=2$. Solid and dashed lines denote polar and equatorial scalings, respectively. 
    The dark green and violet lines in (c) are Eqs.~\ref{eq:Qeta2} and \ref{eq:Qeta3}, respectively, with numerical prefactors. The dotted blue line shows the BG95 model (Eq.~\ref{eq:Qeta4}) with $d_{\eta,2}=\pi/(3\sqrt{2})=0.74$ and $\beta=0$. Parameters: $a=0.35$ in (a,b,d), $\omega=6$ in (a,b,c), and $E_\eta=10^2$ (black) and $E_\eta=10^{-2}$ (orange) in (c,d).}
    \label{fig:Qetafactor}
    \end{center}
\end{figure*}

The second dissipation mechanism considered is Ohmic dissipation, arising from electric currents generated by the interaction of a conducting fluid with a magnetic field. Across the explored parameter space, three regimes emerge as functions of the oscillation-based magnetic Reynolds number $Rm_{\omega}$ and the skin-depth Lundquist number $Lu_\omega$ (Fig.~\ref{fig:Qetafactor}). In the first two regimes ($Lu_\omega<1$), magnetic diffusion dominates.

In the first regime, magnetic perturbations extend through the full core by diffusion, as shown in Fig.~\ref{fig:dissipationdistribution}(c).
$Rm_{\omega}$ is indeed smaller than 1 ($E_\eta\gg1$), meaning that the magnetic skin depth is larger than the radii gap ($L_\eta > (a_\mathrm{m}-a_\mathrm{s})$).
In this configuration, electric currents are simply induced by the inviscid base flow on the imposed magnetic field.
For high magnetic diffusivity, and considering only currents in the outer core, \citet{cebron2025hydromagnetic} have analytically derived the Ohmic dissipation as
\begin{equation}
    \mathcal{D}_\eta = d_{\eta,1} \frac{Le^2(\epsilon\omega)^2}{E_\eta}  \frac{a^3}{1-a^3} = d_{\eta,1}  Lu_\omega^2 \omega(\epsilon\omega)^2 \frac{a^3}{1-a^3}
    \label{eq:ohmicDissipationVolume}
\end{equation}
where $d_{\eta,1}$ is $\pi (1 + 2\omega^{-1})/5$ for the polar mode and $(7/5 + 4 a^3 ) \pi/3$ for the equatorial mode  (see their appendix B.2).
For Earth parameters, these values are approximately 0.8 and 1.6, respectively. 
These prefactors are derived assuming an imposed axial magnetic field, with the polar mode accounting for a rotation-perturbed bulk flow. The lack of the latter for equatorial modes prevents a similar rotation correction (i.e. term in $\omega^{-1}$).
The dissipation in this regime is shown in Figs.~\ref{fig:Qetafactor}(a,c,d), with analytical predictions (black lines) and simulations (black symbols; dashed lines and triangles for equatorial modes). The dot-dashed line in Fig.~\ref{fig:Qetafactor}(c) neglects rotation corrections in $d_{\eta,1}$. Numerical prefactors are 0.7 (polar mode), and 2.4 and 3.9 for prograde and retrograde equatorial modes, with good agreement for the polar case. For polar oscillations, alignment with the magnetic field prevents induced currents in the inner core, whereas equatorial modes generate additional, albeit small, dissipation that may explain the deviations. The corresponding Ohmic quality factor is
\begin{equation}
Q_\eta = \chi d_{\eta,1}^{-1} E_\eta Le^{-2}\omega(1-a^3) = \chi d_{\eta,1}^{-1} Lu_\omega^{-2}(1-a^3) .
\label{eq:Qeta1}
\end{equation}
which scales linearly with $E_\eta$ (right panel of Fig.~\ref{fig:Qetafactor}(a)). Although not relevant for Earth ($Rm_{\omega} \gg 1$, $L_\eta \ll (a_\mathrm{m}-a_\mathrm{s})$), this regime may apply to shallow subsurface oceans with $L_\eta \sim (a_\mathrm{m}-a_\mathrm{s})$. In this case, confinement effects are captured by Eq.~\ref{eq:viscousDissipation}, consistent with analytical predictions (Fig.~\ref{fig:Qetafactor}(c)) and simulations. The frequency dependence (Fig.~\ref{fig:Qetafactor}(d)) follows a quadratic scaling in good agreement with simulations.

The second regime is present at large $Rm_{\omega}$ (small $E_\eta$), and $Lu_\omega$ smaller than 1.
Ohmic dissipation is confined in a skin layer that extends on both sides of the ICB, as shown in Figure~\ref{fig:dissipationdistribution}(d).
We numerically retrieve the following power law for the magnetic dissipation,
\begin{equation}
    \mathcal{D}_\eta = d_{\eta,2} \frac{Le^2(\epsilon\omega)^2}{\sqrt{E_\eta\omega}} \frac{a^2}{(1-a^3)^2} = d_{\eta,2} \frac{Lu_\omega^2(\epsilon\omega)^2}{\sqrt{Rm_\omega}}\omega\frac{a^3}{(1-a^3)^2}  ,
    \label{eq:Deta2}
\end{equation}
where $d_{\eta,2}$ is the prefactor.
The resulting Ohmic quality factor is
\begin{equation}
Q_\eta = \chi d_{\eta,2}^{-1}  Le^{-2}\sqrt{E_\eta\omega}  \omega a(1-a^3)^2 = \chi d_{\eta,2}^{-1} Lu_\omega^{-2}\sqrt{Rm_\omega} (1-a^3)^2 .
\label{eq:Qeta2}
\end{equation}
The linear relationship between $Q_\eta $ and $\sqrt{E_\eta}$ is evident in Fig.~\ref{fig:Qetafactor}(a), left of the vertical dashed line.
The dissipation in this regime is also proportional to the product $Le Lu_\omega$, as shown in the left part of Fig.~\ref{fig:Qetafactor}(b).
This trend is linear up to $Lu_\omega$ close to 1.
The range of $Lu_\omega$ that we can reasonably expect for the Earth is highlighted by the dark green area, in Fig.~\ref{fig:Qetafactor}(b).
The last two regimes are characterized by the quadratic dependence of the magnetic field intensity ($\mathcal{D}_\eta\propto Le^2$), scaled dissipation values characterized by different $Le$ are collapsing on the two master curves in Fig.~\ref{fig:Qetafactor}(a).
This is consistent with the early works of \citet{crossley1974electromagnetic} and \citet{toomre1974diurnal}, which propose a scaling $Q_\eta\propto B_0^{-2}$. Energy is dissipated as Joule heating, $\propto \sigma^{-1}(\nabla \times \bm{B})^2$.
The normalized Ohmic dissipation scales as $D_\eta/Le^2 \propto \sqrt{Rm_\omega}$, i.e. linearly with the magnetic skin depth $L_\eta$, where it is confined.

A third regime is present when $Rm_\omega$ and $Lu_\omega$ are larger than 1.
In this regime, Alfv\'en waves are emitted from the oscillating inner core and propagate along the imposed magnetic field (Fig.~\ref{fig:dissipationdistribution}(e)). The Ohmic dissipation is
\begin{equation}
    \mathcal{D}_{\eta}  = d_{\eta,3} Le (\epsilon\omega)^2 \frac{a^2}{(1-a^3)^2}  =  d_{\eta,3}\frac{Lu_\omega}{\sqrt{Rm_\omega}} \omega (\epsilon\omega)^2  \frac{a^3}{(1-a^3)^2}
\end{equation}
giving the quality factor
\begin{equation}
Q_\eta =  d_{\eta,3}^{-1} \chi Le^{-1}\omega a(1-a^3)^2 =  d_{\eta,3}^{-1}  \chi Lu_\omega^{-1}\sqrt{Rm_\omega}(1-a^3)^2  .
\label{eq:Qeta3}
\end{equation}
In contrast to the first two regimes, dissipation scales linearly with $Le$ (instead of $Le^2$) and is independent of magnetic diffusivity, indicating an Alfv\'en wave drag mechanism. This process also governs the damping of artificial \citep{drell1965drag} and natural satellites, as well as planets in stellar magnetospheres \citep{lai2012dc,bouvier2015protostellar}, where a linear dependence on magnetic field strength and independence from electrical conductivity are likewise observed.

The two large-$Rm_\omega$ regimes can be described analytically by a continuous function of $Lu_\omega$, as first shown by BG95. Using a boundary-layer model, they determined the magnetic field at the inner core surface, computed magnetic tension by surface integration, and estimated $Q_\eta$, restricting the analysis to polar oscillations in an inviscid, unbounded fluid. Despite these simplifications, the BG95 model captures the functional dependence of $Q_\eta$ on $Lu_\omega$, as illustrated by the agreement between the azure dotted line and simulations (orange points) in Fig.~\ref{fig:Qetafactor}(b). The model reproduces well the dissipation of polar modes. Their estimates under Earth conditions are indicated by blue stars, with the upper one corresponding to $B_r=\qty{0.5}{\milli\tesla}$.

Within our notation, BG95 expression for $Q_\eta$ gives
\begin{equation}
    Q_\eta = \frac{1}{2}d_{\eta,2}^{-1} \chi  Lu_\omega^{-2}  \sqrt{Rm_\omega} \left[\mathrm{Im} \left( \frac{1 + \mathrm{i} }{ 1+\sqrt{1-\mathrm{i}Lu_\omega^2/3}} \right) \right]^{-1} (1-a^3)^\beta   ,
\label{eq:Qeta4}
\end{equation}
where the last term in parenthesis is our extension to arbitrary bounding.
In BG95, the prefactor $d_{\eta,2}$ is $\pi/ (3 \sqrt{2}) \approx 0.74$, and $\beta=0$ since they consider an unbounded liquid core.
As illustrated by the azure dotted line in Figure~\ref{fig:Qetafactor}(c), setting $\beta$ to zero results in entirely overlooking the bounding effect.
The model of BG95 was derived assuming a radial magnetic field. To adapt it to our axial magnetic field, we use the root-mean-square value, $B_{r,0}=B_{z,0}/\sqrt{3}$. The equations above take into account this conversion, there $Le$ and $Lu_\omega$ are computed using the axial field.
From the limit of large $Lu_\omega$ of Eq.~\ref{eq:Qeta4}, we can derive the prefactor of the third regime:  $d_{\eta,3} = 2\sqrt{6} d_{\eta,2}$.
For thorough consistency with the approach of BG95, which considers only the magnetic tension effects, the exponent $\beta$ should equal 1  to account for bounding effects.

\citet{cebron2025hydromagnetic} extended the BG95 boundary layer model to bounded domains, including viscous and rotational effects. For small $Pm$, viscous contributions are negligible and not considered further. Their analysis also covers equatorial oscillations and goes beyond the magnetic tension approximation. Ohmic dissipation is computed from volume integration of radial derivatives of the analytically predicted tangential magnetic field, following \citet{batchelor1967introduction} (Chapter 5). For a bounded, inviscid fluid with an imposed radial field, the solution retains the form of Eq.~\ref{eq:Qeta4} with $\beta=2$. This quadratic scaling with fluid volume is confirmed by our simulations (Supplementary Material S4). The corresponding prefactors are $d^{\star}_{\eta,2} = \pi \sqrt{2}/4 \approx 1.11$ for polar modes and $d^\star_{\eta,2} = \pi \sqrt{2}/2 \approx 2.22$ for equatorial modes (orange curves in Fig.~\ref{fig:Qetafactor}), the latter being twice the former when rotation corrections are neglected.

Our simulations (at $\omega=6$) are best described by Eq.~\ref{eq:Qeta4} with $d_{\eta,2}=0.73$ for the polar mode, and 2.5 and 3.2 for prograde and retrograde equatorial modes. These prefactors are used in Fig.~\ref{fig:Qetafactor}(b) (dark green and violet lines for Eqs.~\ref{eq:Qeta2} and \ref{eq:Qeta3}, respectively). The equatorial prefactors agree well with analytical predictions. Equation~\ref{eq:Qeta2} reproduces effectively the dependence on radii ratio (Fig.~\ref{fig:Qetafactor}(c)). Confinement effects are significant, increasing Ohmic dissipation by several orders of magnitude in thin shells.

In all three regimes, $Q_\eta$ is independent of the oscillation velocity $\epsilon\omega$ due to its small amplitude. It is also largely independent of the spin rate, aside from minor prefactor corrections; the second terms in Eqs.~\ref{eq:Qeta1}–\ref{eq:Qeta3} are $\Omega_\mathrm{o}$-independent. Our scaling further shows that $\mathcal{D}_\eta$ and $Q_\eta$ are independent of fluid viscosity, consistent with the low-$Pm$ regime of the outer core \citep{cebron2025hydromagnetic}. The inviscid approximation of BG95 thus captures the relevant physics. To explore smaller $E_\eta$ under constraints on $E$, we increased $Pm$ up to 0.1, obtaining a good collapse of results at fixed $E_\eta$ (Fig.~\ref{fig:Qetafactor}(b)). Having established the scaling laws and prefactors for both dissipation mechanisms, we now provide estimates for Earth conditions.

\section{Discussion and conclusion} \label{sec:discCON}
Slichter modes have so far eluded direct detection. This absence of verified observations can be attributed to one or more of the following factors:
(1) sufficiently strong excitation events able to produce detectable gravity perturbations are rare;
(2) the current estimation of the oscillation period may be inaccurate;
(3) the energy injected into the system is dissipated rapidly.
The first point (1) seems consistent with the fact that an earthquake of magnitude 9.7 is necessary to produce a detectable gravity signal $\sim$ \qty{e-11}{\meter\per\square\second}, as discussed by \citet{rosat2011review}.

After five decades of study, it is clear that the oscillation period (2) is mainly controlled by gravitational forces.
Other phenomena such as solid layer elasticity and ellipticity, liquid compressibility, stratification, and phase change introduce small or even negligible corrections \citep{wu1994gravity}.
In the supplementary material S1, we provide a clarification for why the impact of phase change is insignificant.
Ultimately, the densities of the layers are the main contributors to the inaccuracies in the period estimate.
The emphasis of this study is on the third aspect (3), because the dissipation mechanisms were previously poorly constrained.
In this work, we have performed simulations to compute scaling laws for the viscous and Ohmic dissipation for both the polar and equatorial modes.

\subsection{Viscous dissipation}
The limited magnitude of the oscillations confines them within the viscous boundary layer.
The scaling shown in Figure~\ref{fig:Qnufactor} is consistent over 7 orders of magnitude of $E$, therefore, we can safely extrapolate down to the Earth's conditions.
Table~\ref{tab:results}, summarizes the estimates of the viscous dissipation, quality factor, and decay time for the three modes in the Earth's condition.
The values are obtained by substituting Earth's parameter (Tab.~\ref{tab:parameters}) and the numerical prefactors into equations~\ref{eq:viscousDissipation} and \ref{eq:viscousQualityFactor}.
Estimates are at the frequencies reported in Table 1 of \citet{rieutord2002slichter}, see \ref{sec:Earthparameters}.
\begin{table*}
    \centering
    \begin{tabular}{lc|ccc|ccc}
    \hline
    mode   &  $\omega$  & $D_\nu$ & $Q_\nu$ & $\tau_\nu$ & $D_\eta$ & $Q_\eta$ & $\tau_\eta$ \\
    unit   &    & \si{\watt} &  & y  & \si{\watt} &  & y \\ \hline
    polar   &  5.73   & 12 & \num{7.5e6}& \num{1.1e3} & \num{8.5e2} & \num{e5}& \num{16}\\
    eq. prograde     & 5.20   & 20 &\num{3.4e6} & \num{5.7e2} & \num{2.4e3} & \num{2.9e4}& \num{4.8}   \\
    eq. retrograde   &  6.26  & 27 & \num{4.2e6} & \num{5.9e2} & \num{4e3} & \num{2.8e4}& \num{4}\\
   \hline
    \end{tabular}
    \caption{Viscous and Ohmic dissipation, quality factors and decay times. Values of Table~\ref{tab:parameters}.}
    \label{tab:results}
\end{table*}
Due to the extremely low value of the viscosity of the outer core, the role of viscous dissipation is negligible.
This also results in the impossibility of using the damping rate of the detected Slichter modes to estimate the outer core viscosity.
In the absence of a magnetic field, the translational oscillation can theoretically persist for up to a few hundred years. This finding may be relevant for celestial bodies such as planets or moons that lack magnetic fields but possess liquid interiors.

\subsection{Ohmic dissipation}
The Ohmic quality factor, dissipation, and decay time for the three modes at the Earth's conditions are summarized in Table \ref{tab:results}.
These values are calculated using the numerical prefactor and Earth's parameters (see Tab.~\ref{tab:parameters}) into Eqs.~\ref{eq:Deta2} and~\ref{eq:Qeta2}.
These results are in the range estimated by the semi-analytical computation of \citet{buffett1995magnetic}, with decay times spanning from 150 days to 2200 years.
For the specific values adopted in this work, the estimated decay time is around 16 years for the polar mode, and 3 and 4 years for the retrograde and prograde equatorial modes, respectively.
These values remain only indicative of the order of magnitude, due to the rather large uncertainties, especially in the magnetic field intensity and electrical conductivity at the ICB.

If the 420-day anelastic damping time estimated by \citet{crossley1991excitation} is confirmed, this mechanism would dominate over Ohmic dissipation, becoming the leading cause of damping.
In this respect, a valuable extension of this work would be the study of the inner core’s rheological response. Nevertheless, since no dissipation process can quickly dampen the translational modes, it is possible to witness their existence for extended periods of time. Their non-detection is more plausibly attributed to weak excitation rather than strong damping.

\subsection{Beyond the Earth}
Translational oscillations are not unique to Earth’s inner core and have been studied in icy moons and planets such as Mercury \citep{escapa2011free,coyette2014slichter,grinfeld2005motion,coyette2012period}. These studies provide period estimates and assess detectability, typically using unbounded dissipation models \cite[e.g.][]{buffett1995magnetic}. However, in shallow subsurface oceans, confinement effects become significant and can strongly enhance dissipation (Fig.~\ref{fig:Dnuradii}).

Future work should address the role of stable stratification, neglected here but potentially important \cite[e.g. in Mercury, see][]{seuren2023effects}, on both periods and damping. These works provide period estimates and discuss their potential detectability by space missions. When dissipation is computed, unbounded models like \cite{buffett1995magnetic} are typically employed.
Ultimately, if advancements in measurement techniques enable the detection of these oscillations in the Earth or other planetary bodies, the proposed correlations could significantly enhance our ability to constrain the interior properties of these objects, offering a novel probe into the dynamics of deep planetary layers.

\section*{Acknowledgement}
This work benefits from the funding from the ERC under the European Union Horizon 2020 research and innovation program via GRACEFUL Grant No. 855677 (for PP \& MM), and THEIA Grant No. 847433 (for PP \& DC). Computations on the GRICAD infrastructure (\url{https://gricad.univ-grenoble-alpes.fr}), and HPC resources (Jean Zay V100 and H100) of IDRIS under allocation AD010413621 and A0160407382 attributed by GENCI (Grand Equipement National de Calcul Intensif). 

The supporting data will be made available upon publication and can also be provided upon request.
Numerical simulations were performed using the XSHELLS code, which is freely available at \url{https://nschaeff.bitbucket.io/xshells/}.

\bibliographystyle{elsarticle-harv}
\bibliography{ref.bib}

\newpage
\setcounter{page}{1}
\section*{SUPPLEMENTARY MATERIAL}
\renewcommand{\thefigure}{S\arabic{figure}}   
\setcounter{figure}{0} 
\renewcommand{\thesection}{S\arabic{section}}
\setcounter{section}{0} 
\renewcommand{\appendixname}{}

\section*{Introduction}
The supporting information comprises supplementary texts and figures. Specifically:
\textbf{S1} provides a detailed discussion on the effect of phase change in the presence of translational oscillations.
\textbf{S2} offers a comparison between the pseudo-spectral code and a finite element code to validate our approach.
\textbf{S3} briefly addresses the limitations of our approximated boundary condition used for modeling translational oscillations.
\textbf{S4} addresses the role of bounding for the viscous and magnetic dissipation.
\textbf{S5} discusses the assumptions and validity limits in terms of inner core viscosity.

\section*{Text S1. Effect of phase change at the solid oscillating boundary}\label{A1:phase_change}
\begin{figure}
	\centering
	\includegraphics[width=0.7\linewidth]{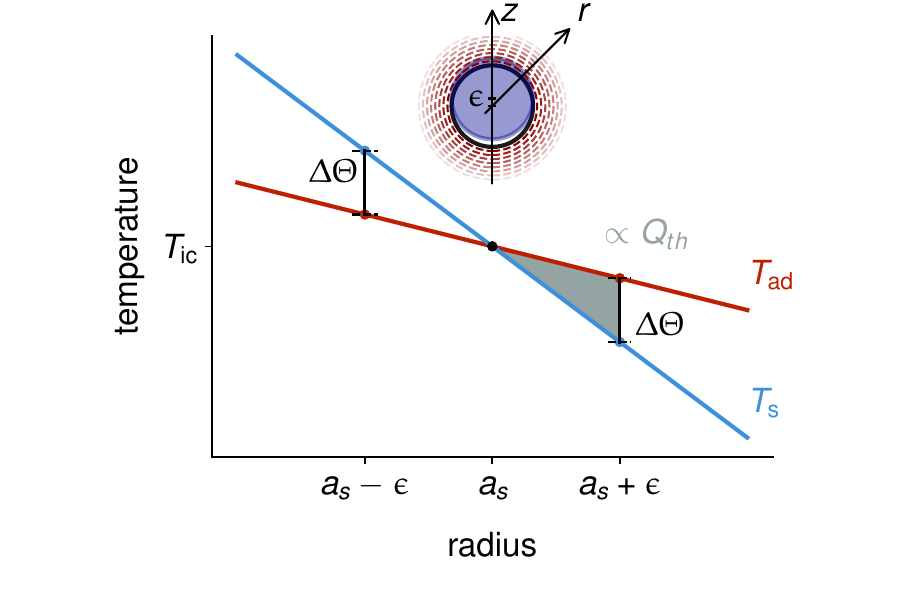}
	\caption{Temperature profiles near the inner core boundary: the outer core adiabat is shown in red, and the solidification temperature in blue. The top panel schematically illustrates inner core oscillations within this temperature field.}
	\label{fig:phase_chage}
\end{figure}

We estimate the effect of phase change on translational oscillations, extending the analysis of \citet{wu1994gravity}. We consider the polar mode, noting that the results also apply to equatorial modes. An instantaneous thermodynamic response is assumed, with immediate melting or solidification, providing a conservative upper bound; kinetic effects would further reduce this contribution. The outer core follows an adiabatic temperature gradient (red line in Fig.~\ref{fig:phase_chage}), defined as
\begin{eqnarray}
	\frac{\partial T_\mathrm{ad}}{\partial r} = \frac{\alpha g T}{c_p} &=& \frac{\SI{1.1e-5}{\per\kelvin}\cdot\SI{4.4}{\meter\per\square\second}\cdot \SI{5.6e3}{\kelvin}}{\SI{800}{\joule\per\kilogram\per\kelvin}} \nonumber \\ &=& \SI{3.4e-4}{\kelvin\per\meter}
\end{eqnarray}
in which $\alpha$ and $c_p$ are the thermal expansion coefficient and the specific heat capacity of the outer core fluid, $g$ and $T$ are the gravity acceleration and the absolute temperature close to the inner core boundary (ICB), the numerical values are from \citet{deguen2013thermal}.
If we assume an inner core displacement $\epsilon_\mathrm{s}$ of \SI{10}{\centi\meter}, the temperature variation experienced is in the order of \SI{3e-5}{\kelvin}.
The inner core boundary is a solidification front; the solidification isotherm gradient can be expressed as
\begin{eqnarray}
	\frac{\partial T_s}{\partial r} = \frac{\partial T_s}{\partial p} \rho_l g &=& \SI{e-8}{\kelvin\per\pascal}\cdot\SI{1.2 8e4}{\kilogram\per\cubic\meter}\cdot\SI{4.4}{\meter\per\square\second} \nonumber \\ &=& \SI{5.6e-4}{\kelvin\per\meter}
\end{eqnarray}
where $\frac{\partial T_s}{\partial p} $ is the Clapeyron slope, $\rho_l$ is the liquid core density at the ICB \citet{deguen2013thermal}. The solidification
isotherm is shown in blue in Fig.~\ref{fig:phase_chage}.
The difference between the two gradients is the under-cooling gradient
\begin{equation}
	\frac{\partial \Delta \Theta}{\partial r} = \left| \frac{\partial T_\mathrm{s}}{\partial r} - \frac{\partial T_\mathrm{ad}}{\partial r} \right| = \SI{2.2e-4}{\kelvin\per\meter} \ .
\end{equation}

\noindent The undercooling experienced by the translated inner core ($\epsilon_\mathrm{s} =~$ \SI{10}{\centi\meter}) can be estimated to be in the order of \SI{2e-5}{\kelvin}.
This value improves the one proposed by \citet{wu1994gravity}, which got a comparable value assuming a \qty{1}{\meter} displacement.
In the following, we estimate the percentage of the displaced inner core that is solidified.
This can be computed with a simple energy balance; we assume that the amount of thermal energy ideally exchanged during the inner core translation is balanced by the latent heat.
The amount of thermal energy ideally exchanged can be estimated as
\begin{equation}
	\mathcal{Q}_\mathrm{th} = \rho c_p \Delta \Theta \Delta V_\mathrm{osc} \approx \frac{1}{2}\rho c_p \left|\frac{\partial T_\mathrm{s}}{\partial r} - \frac{\partial T_\mathrm{ad}}{\partial r}  \right| \epsilon_\mathrm{s} A \epsilon_\mathrm{s} ,
\end{equation}
where the volume displaced is approximated as $A \epsilon_\mathrm{s}$ (in which $A$ is a geometric area). The amount of latent heat can be computed as
\begin{equation}
	\mathcal{Q}_\mathrm{pc} =  \rho L \Delta V_\mathrm{pc}  \approx \rho L \delta_\mathrm{pc} A
\end{equation}
in which $A \delta_\mathrm{pc}$ is the approximated volume affected by the phase change.
The percentage of the displaced inner core that is solidified is
\begin{equation}
	\frac{\delta_\mathrm{pc}}{\epsilon_\mathrm{s}} = \frac{1}{2} \frac{c_p}{L}  \left|\frac{\partial T_\mathrm{s}}{\partial r} - \frac{\partial T_\mathrm{ad}}{\partial r}  \right|\epsilon_\mathrm{s} = \SI{e-7}{} \epsilon_\mathrm{s} \ .
\end{equation}
If we consider a \SI{10}{\centi\meter} displacement, the resulting phase change thickness is \SI{1}{\nano\meter}, showing the insignificance of phase change in this process.

\begin{figure}
	\centering
	\subfigure{
    \includegraphics[width=0.45\linewidth]{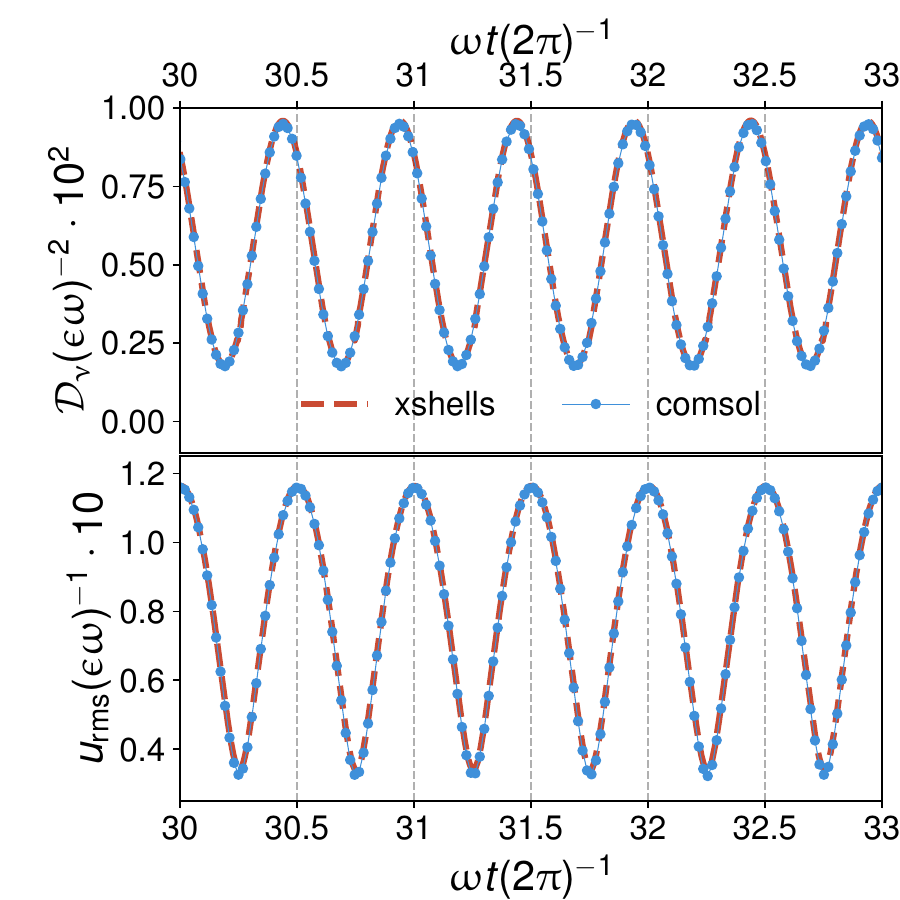}
    }
    \subfigure{
    \includegraphics[width=0.45\linewidth]{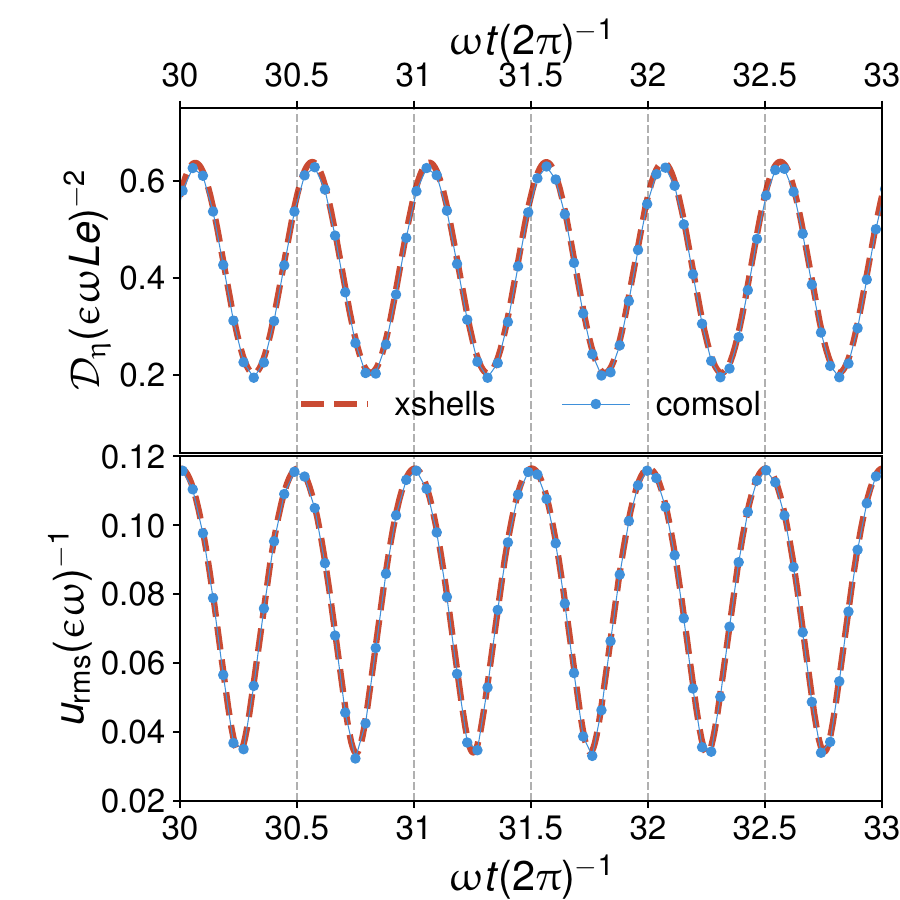}
    }
    \label{fig:Detaradii}
	\caption{Dissipation and rms-velocity $u_{\text{rms}} = \sqrt{\langle u^2\rangle_V}$ vs. time, from XSHELLS and COMSOL. (left) Viscous dissipation at $E=10^{-4}$, $a=0.35$, $\omega=6$,$\epsilon\omega=0.01225$. (right) Ohmic dissipation at $Le=2\cdot10^{-4}$, $Pm=10^{-2}$. $\langle\rangle_V$ being volume average.}
	\label{fig:timeEvolutionViscous}
\end{figure}

\section*{Text S2. Numerical comparison among different methods}
\begin{figure}
	\centering
		\subfigure{
	        \includegraphics[width=0.42\linewidth]{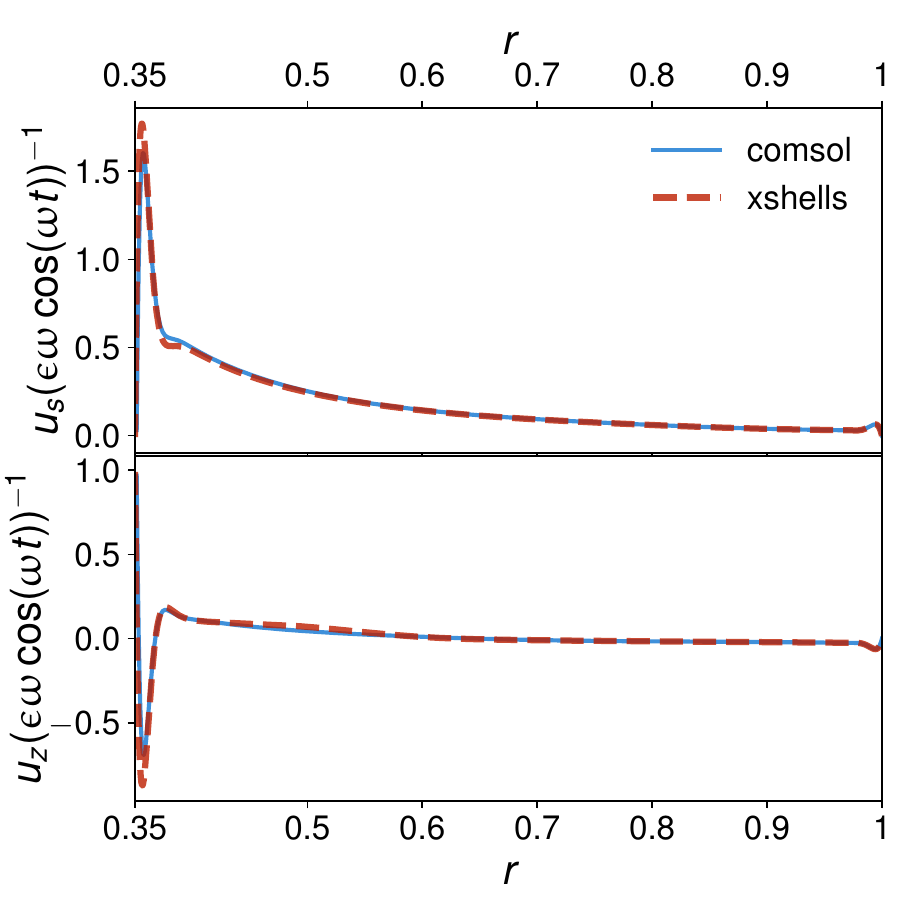}
	    }
	    \subfigure{
	    		\includegraphics[width=0.53\linewidth]{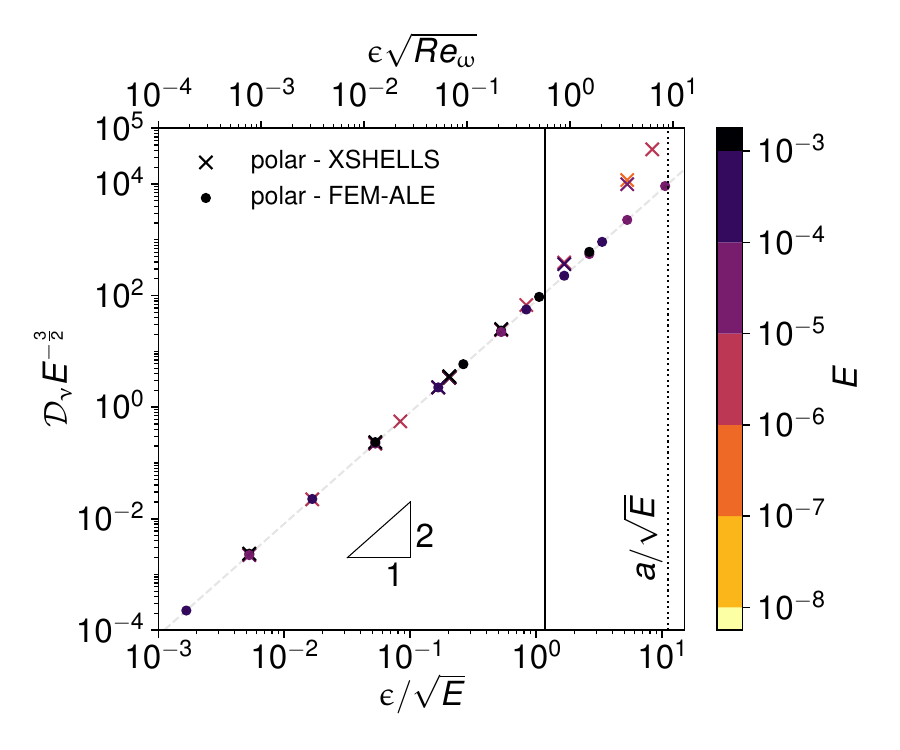}
	    }
	\caption{(left) Velocity profile for the polar mode with no-slip boundary conditions ($E=10^{-4}$, $a=0.35$, $\omega=6$, $\epsilon\omega=0.01225$). (right) Scaled viscous dissipation vs. oscillation amplitude from XSHELLS (fixed boundary with approximated conditions) and COMSOL-ALE (moving inner boundary), at $a=0.35$, $\omega=6$. }
	\label{fig:profiles}
\end{figure}

To validate our results, we compare the pseudo-spectral finite-difference solver XSHELLS with the finite-element solver COMSOL Multiphysics\textsuperscript{\textregistered}. Both hydrodynamic and magnetohydrodynamic cases are considered. COMSOL implements the inner core motion using an arbitrary Lagrangian–Eulerian method that deforms the mesh.  For small amplitude oscillations, the two approaches show strong agreement in integral quantities, including dissipation and root-mean-square velocity (Fig.~\ref{fig:timeEvolutionViscous}), as well as in local velocity profiles (Fig.~\ref{fig:profiles}a), for both cylindrical radial and axial components at co-latitude $\vartheta=\qty{45}{\degree}$ and azimuth $\varphi=\qty{0}{\degree}$.

\section*{Text S3. Limits of validity of the approximated boundary condition}
To model inner core translational oscillations, XSHELLS imposes the boundary conditions (3)–(4), which approximate the true geometric displacement. We assess this approximation using simulations with explicit inner core motion, implemented via the arbitrary Lagrangian–Eulerian (ALE) method in COMSOL. As shown in Fig.~\ref{fig:profiles}(b), both approaches agree for oscillation amplitudes within the skin layer depth: the approximated results (‘x’) closely match the ALE solutions (dots). At larger amplitudes, the approximation overestimates dissipation, although viscous dissipation retains the same scaling. In the following, we adopt the approximated boundary condition, valid within the skin layer regime.

\section*{Text S4. Bounding effects on dissipation mechanism}
\begin{figure}
	\centering
	\subfigure{
    \includegraphics[width=0.47\linewidth]{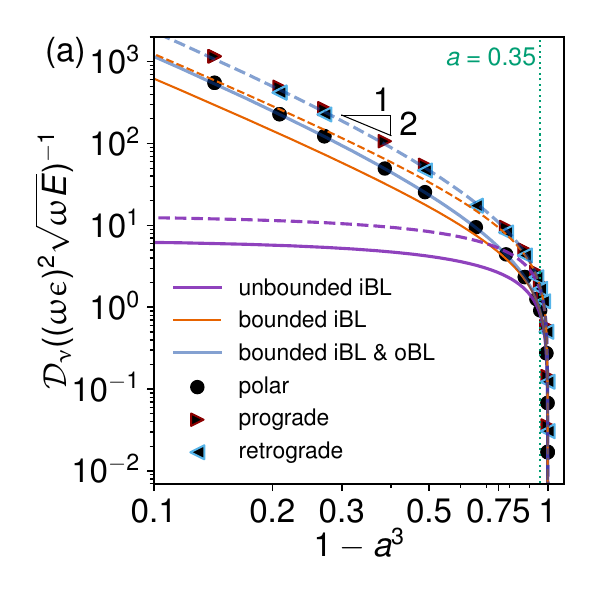}
    }
    \subfigure{
    \includegraphics[width=0.47\linewidth]{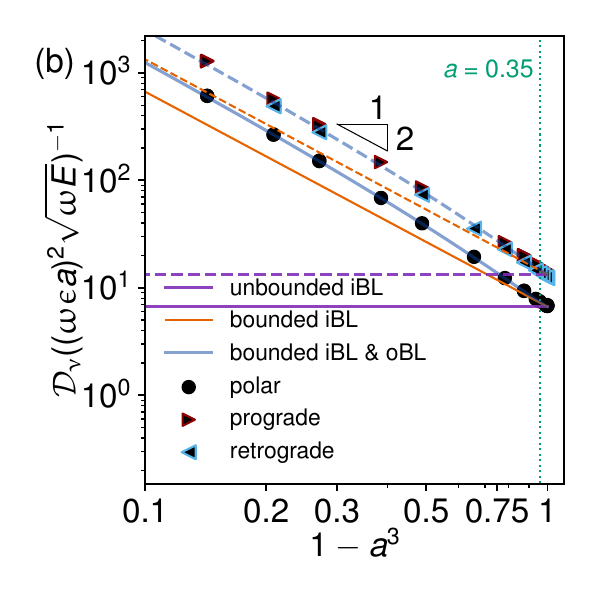}
    }
	\caption{Rescaled viscous dissipation vs. fluid volume $1-a^3$ at $E$=\qty{e-8}, $\omega$=6.}
	\label{fig:scalingViscousFluidVolume}
\end{figure}

Figure \ref{fig:scalingViscousFluidVolume}(a,b) shows that viscous dissipation scales quadratically with the inverse fluid volume $(1 - a^{3})^{-1}$. Normalizing by the geometric factor $a^2$ highlights this trend (Fig.~\ref{fig:scalingViscousFluidVolume}(b)). In shallow layers (small $1-a^3$), dissipation from the outer boundary layer, scaling as $a^4$, becomes significant (light blue line). Overall, the derived prefactor accurately captures confinement effects for both polar and equatorial modes.

\begin{figure}
	\centering
	\subfigure{
    \includegraphics[width=1.02\textwidth]{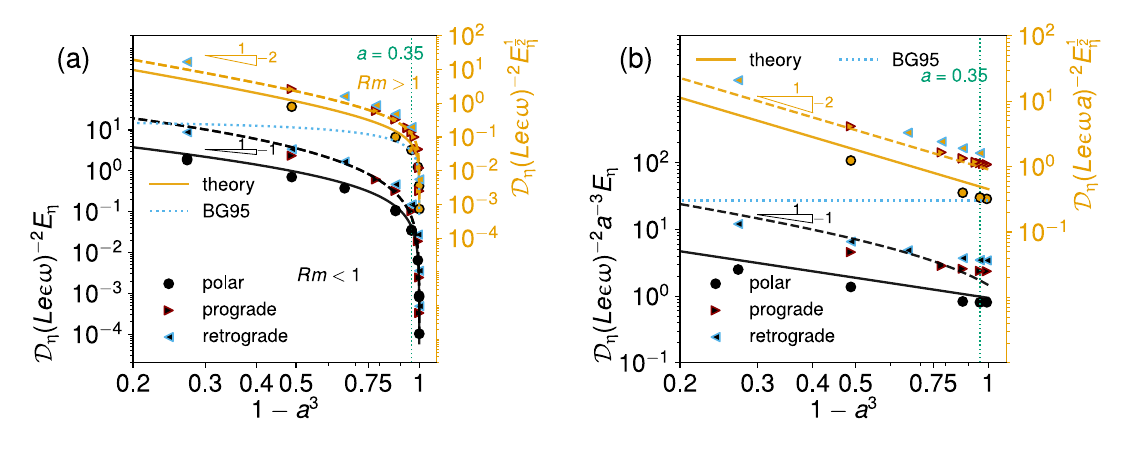}
    }
	\caption{Rescaled Ohmic dissipation vs. fluid volume ($1-a^3$). Parameters: black points: $E=10^{-4}$, $\omega=6$, $Pm=10^{-6}$, orange points: $E=10^{-8}$, $\omega=6$, $Pm=10^{-6}$.
	}
	\label{fig:scalingOhmicFluidVolume}
\end{figure}

Figure \ref{fig:scalingOhmicFluidVolume}(a,b) shows that Ohmic dissipation scales with $(1 - a^{3})^{-1}$, following the expected linear law in the diffusive regime (black points) and the expected quadratic law in the skin-layer regime (orange points). By comparing with numerical results, we conclude that the $a^3$ term in the analytical prefactor $d_{\eta,1}$  for equatorial mode dissipation is likely compensated by an analogous term in the inner core.

\section*{Text S5. Assumptions on the inner core rheology}
The viscosity of Earth’s inner core remains poorly constrained, spanning several orders of magnitude. Estimates range from $2$–\SI{7e14}{\pascal\second} based on nutation studies \citep{koot2011viscosity} to an upper bound of $\sim$\SI{3e17}{\pascal\second} from core–mantle coupling \citep{davies2014the}, while mineral physics suggests \SIrange{e16}{e18}{\pascal\second} for hcp iron \citep{ritterbex2020viscosity}. More recent estimates propose \SIrange{e14}{2e16}{\pascal\second} \citep{xu2025viscosities}. We neglect viscous deformation of the inner core and assess this assumption by comparing the oscillation period to the stress relaxation time $\tau = \rho_\mathrm{s} \nu_\mathrm{s}/E_\mathrm{s}$ from a Maxwell model, where $E_\mathrm{s} \approx \SI{e12}{\pascal}$ (PREM; \citealp{dziewonski1981preliminary}). The inner core behaves elastically at the oscillation timescale for viscosities above \SI{2e16}{\pascal\second}, consistent with mineral physics \citep[e.g.][]{ritterbex2020viscosity} and geodynamical estimates \citep{davies2014the}.

\end{document}